\documentclass[a4paper]{jpconf}

\usepackage{amsmath,amssymb,hyperref,times,graphicx,bm,bbold,cite}
\def\<{\langle}
\def\>{\rangle}

\begin{document}

\bibliographystyle{iopart-num}
\title{Surface critical behavior of the three-dimensional $O(3)$ model}
\author{F Parisen Toldin}
\address{Institut f\"ur Theoretische Physik und Astrophysik, Universit\"at W\"urzburg, Am Hubland, D-97074 W\"urzburg, Germany}
\ead{francesco.parisentoldin@physik.uni-wuerzburg.de}
\begin{abstract}
  We report results of high-precision Monte Carlo simulations of a three-dimensional lattice model in the $O(3)$ universality class, in the presence of a surface. By a finite-size scaling analysis we have proven the existence of a special surface transition, computed the associated critical exponents, and shown the presence of an extraordinary phase with logarithmically decaying correlations.
\end{abstract}

\section{Introduction}
\label{sec:intro}
A system
in the vicinity of a critical point exhibits a variety of interesting features, such as power-law singularities and scaling behavior in many observables.
One of the most fascinating aspects is the emergence of universality: critical exponents associated to the aforementioned singularities, and other quantities, are independent of the local details of interactions. They are rather determined by the global features of the system, such as the symmetry group, the pattern of symmetry-breaking, dimensionality and range of interactions.
The theory of Renormalization Group (RG) provides a framework for understanding and predicting the emergence of universality through a suitably defined flow of Hamiltonians, the fixed points of which control the critical behavior and define the so-called Universality Classes (UCs) \cite{Cardy-book}.

While the singular behavior associated to the onset of a phase transition occurs, in principle, in the thermodynamic limit only, real physical systems naturally have boundaries.
Their presence is the source of rich phase diagrams and critical behavior, that has been the target of many experimental \cite{Dosch-book} and theoretical \cite{Binder-83,Diehl-86,Pleimling-review} investigations.
According to RG theory, a given {\it bulk} fixed point controlling the critical behavior in the thermodynamic limit potentially splits into several fixed points associated to the critical behavior on the boundary \cite{Cardy-book,Diehl-86}, thereby defining surface or, more generally, boundary UCs.
This implies that critical exponents and other universal quantities on the boundary differ from bulk ones. Furthermore, for a given system at a critical point the boundary may exhibit diverse critical behavior, depending on the strength of boundary interactions.
Surface UCs also determine the critical Casimir force \cite{FG-78,Krech-94,Krech-99,Gambassi-09,GD-11,MD-18}.

\begin{figure}
  \includegraphics[width=0.47\linewidth]{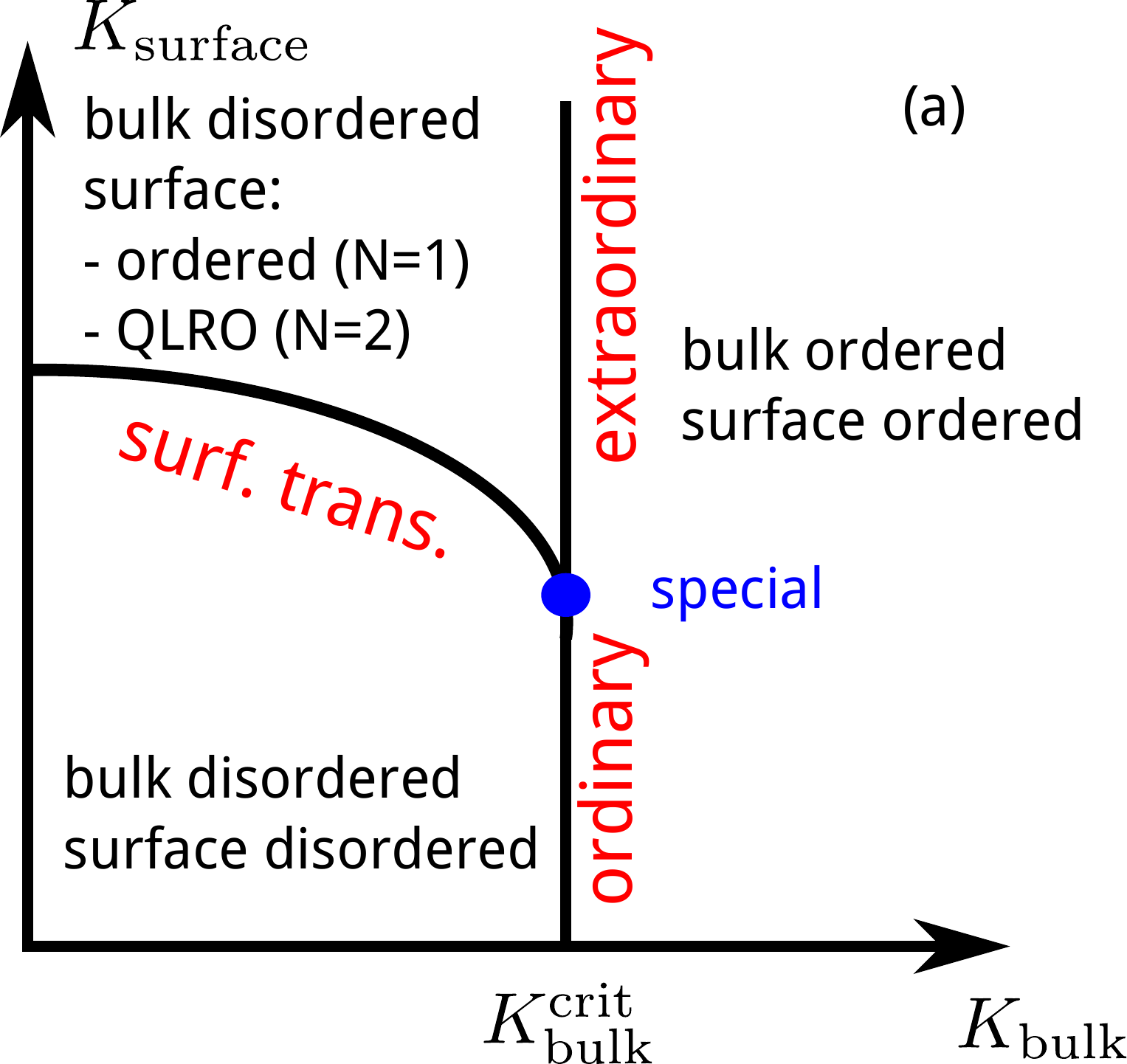}
  \hspace{0.04\linewidth}
  \includegraphics[width=0.47\linewidth]{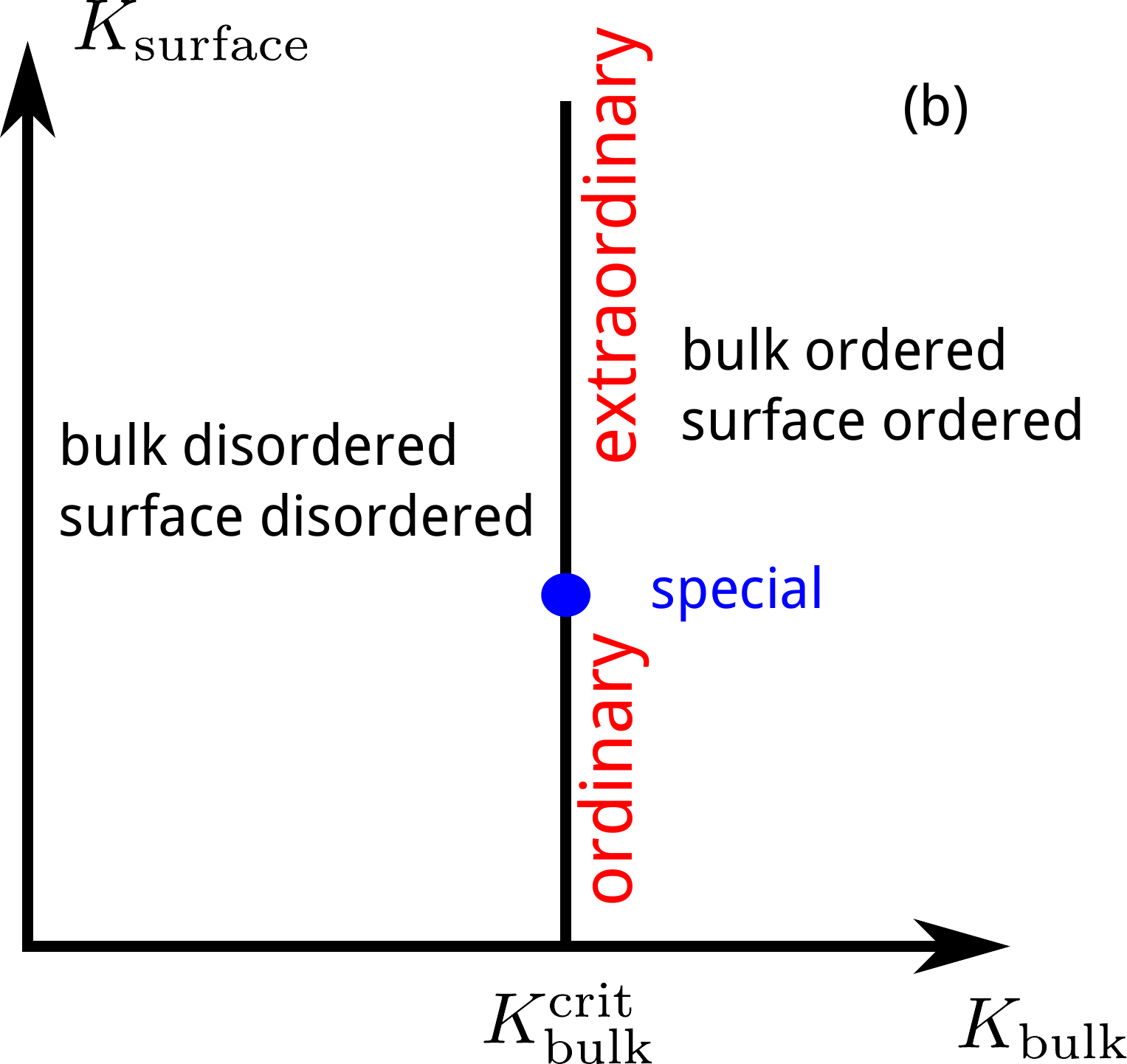}
  \caption{Bulk-surface phase diagram of the three-dimensional O(N) model, bounded by a 2D surface, for $N=1,2$ (a) and $N=3$ (b).}
  \label{phasediag}
\end{figure}

The simplest setup where this physics is realized is the case of a semi-infinite $d{-}$dimensional system bounded by a $(d{-}1){-}$dimensional surface.
In this context, due to its physical relevance, the critical behavior of the classical 3D $O(N)$ model represents one of the most significant UC \cite{PV-02}.
In Fig.~\ref{phasediag} we sketch the bulk-surface phase diagram, as a function of the bulk and surface coupling constants. For $N=1,2$ a surface transition line, in the presence of a disordered bulk, separates a disordered surface from an ordered one, for $N=1$, or from a surface possessing quasi-long range order (QLRO), for $N=2$.
For a critical bulk, and as a function of the surface coupling, we distinguish an ordinary and extraordinary transition lines. Surface, ordinary and extraordinary lines meet at a multicritical point, the so-called special UC \cite{Binder-83,Diehl-86}.
For $N=3$, no surface transition exists \cite{PV-02}, hence the phase diagram topology does not necessarily mandate a special point.
While early Monte Carlo (MC) studies supported the absence of a special point \cite{Krech-00}, a later MC investigation reported a possible Berezinskii-Kosterlitz-Thouless-like surface transition \cite{DBN-05}.
More recently, the problem has received a renewed attention in the context of conformal
field-theory approaches \cite{MO-95,LRVR-13,GLMR-15,BGLM-16,LM-17,LMT-18,MRZ-19,KP-20,DHS-20,BDPLVR-20,Metlitski-20,GGLVV-21,PKMGM-21},
classical \cite{PT-20,HDL-21} and
quantum critical behavior \cite{SS-12,ZW-17,DZG-18,WPTW-18,WW-19,JXWX-20,ZDZG-20,WW-20,DZGZ-21}.
In particular, quantum MC investigations have focused on dimerized spin-$1/2$ \cite{SS-12,ZW-17,DZG-18,WPTW-18} and spin-$1$ \cite{WW-19} systems in $d=2$, which exhibit a second-order quantum phase transition in the classical 3D $O(3)$ UC.
In the presence of an edge, these models have been shown to display ordinary, as well as nonordinary boundary exponents, depending on the geometrical setup.
A recent field-theoretical study has predicted the existence of a so-called ``extraordinary-log'' phase at the surface of a critical 3D $O(N)$ model.
This phase exists for $N < N_c$, with $N_c>2$ \footnote{We remark that $N_c$ does not need to be an integer.}, and
is characterized by surface correlations decaying as a power of a logarithm. The associated exponent is universal, and it is determined by some amplitudes of the normal UC \cite{Metlitski-20}.
This is realized by applying a symmetry-breaking field on the boundary \cite{Binder-83,Diehl-86,BM-77,BC-87}.

Motivated by these advancements,
in Ref.~\cite{PT-20} we have investigated the classical surface $O(3)$ UC by means of MC simulations of an improved lattice model, where leading scaling corrections are suppressed.
A finite-size scaling (FSS) analysis has shown the existence of a special transition and of an extraordinary phase consistent with the extraordinary-log scenario of Ref.~\cite{Metlitski-20}.
In the following we summarize the results of Ref.~\cite{PT-20}.

\section{Model}
\label{sec:model}
We have simulated the $\phi^4$ model on a three-dimensional lattice of size $L$ in all directions, applying periodic boundary conditions (BCs) along two directions, and open BCs on the remaining one.
The reduced Hamiltonian ${\cal H}$, such that the Gibbs weight is $\exp(-\cal H)$, is
\begin{equation}
    {\cal H} = -\beta\sum_{\< \vec{\imath}\ \vec{\jmath}\ \>}\vec{\phi}_{\vec{\imath}}\cdot\vec{\phi}_{\vec{\jmath}}
    -\beta_{s,\downarrow}\sum_{\< \vec{\imath}\ \vec{\jmath}\ \>_{s\downarrow}}\vec{\phi}_{\vec{\imath}}\cdot\vec{\phi}_{\vec{\jmath}}
    -\beta_{s,\uparrow}\sum_{\< \vec{\imath}\ \vec{\jmath}\ \>_{s\uparrow}}\vec{\phi}_{\vec{\imath}}\cdot\vec{\phi}_{\vec{\jmath}}
    +\sum_{\vec{\imath}}[\vec{\phi}_{\vec{\imath}}^{\,2}+\lambda(\vec{\phi}_{\vec{\imath}}^{\,2}-1)^2],
  \label{model}
\end{equation}
where $\vec{\phi}_{\vec{x}}$ is a three-components real field on the lattice site $\vec{x}=(x_1,x_2,x_3)$, indicated as a 3D vector.
In Eq.~(\ref{model}) the first sum extends over nearest-neighbor pairs of sites where at least one belongs to the inner bulk, the second and third sum extend to lattice sites on the lower and upper surface, and the last sum is over all lattice sites.
The coupling constants $\beta$ and $\lambda$ determine the bulk critical behavior.
In the $(\beta,\lambda)$ plane the bulk displays a line of continuous phase transitions in the $O(3)$ UC \cite{CHPRV-02,PV-02}, and in the limit $\lambda\rightarrow\infty$ the model reduces to the standard $O(3)$ hard spin model.
At $\lambda=5.17(11)$ the model is {\it improved} \cite{Hasenbusch-20}, i.e., leading bulk scaling corrections $\propto L^{-\omega}$, $\omega=0.759(2)$, are suppressed. Next-to-leading scaling corrections due to a nonrotationally invariant irrelevant operator have an exponent $\omega_\text{nr}\approx 2$ \cite{Hasenbusch-20}, hence they decay very fast.
Improved lattice models are particularly useful in high-precision numerical studies of critical phenomena \cite{PV-02}, in particular for boundary critical phenomena \cite{Hasenbusch-09b,Hasenbusch-10c,PTD-10,Hasenbusch-11,Hasenbusch-11b,Hasenbusch-12,PTTD-13,PT-13,PTTD-14,PTAW-17,PT-20} because they allow a better control of scaling corrections.
In our MC simulations we have fixed $\lambda=5.2$ and $\beta=0.68798521$, for which the model is critical \cite{Hasenbusch-20}.
The coupling constants $\beta_{s,\downarrow}$, $\beta_{s,\uparrow}$ in Eq.~(\ref{model}) control the strength of boundary interactions.
To study the surface critical behavior we have set $\beta_{s,\downarrow}=\beta_{s,\uparrow}=\beta_s$ and examined various surface observables as a function of $\beta_s$.
MC simulations have been performed by combining Metropolis, overrelaxation, and Wolff single-cluster updates \cite{Wolff-89,PT-20}.

\section{Results}
\label{sec:results}
\subsection{Special transition}
\label{sec:results:special}
A standard method to locate the onset of a continuous phase transition consists in a FSS analysis of RG-invariant observables \cite{Privman-90,PV-02}.
To study the special UC we have proceeded in two steps.
First, we have studied an RG-invariant quantity, determining its critical-point value.
Subsequently, we have employed this value in a FSS analysis of other observables, to compute the critical exponents at the special transition.

According to RG, close to a surface phase transition at $\beta_s=\beta_{s,c}$, an RG-invariant observable $R$ behaves as
\begin{equation}
  R = f((\beta_s-\beta_{s,c})L^{y_{\rm sp}}),
  \label{FSS_RGinv}
\end{equation}
where $y_{\rm sp}$ is the scaling dimension of the relevant scaling field associated with the transition and we have for the moment neglected scaling corrections.
We have analyzed the surface Binder ratio $U_4$, defined as
\begin{equation}
  U_4 \equiv \frac{\<(\vec{M}_s^2)^2\>}{\<\vec{M}_s^2\>^2}, \qquad \vec{M}_s\equiv \sum_{\vec{\imath}\in\text{surface}}\vec{\phi}_{\vec{\imath}}.
  \label{U4}
\end{equation}

\begin{figure}
  \begin{minipage}{0.45\linewidth}
    \centering
    \includegraphics[width=\linewidth]{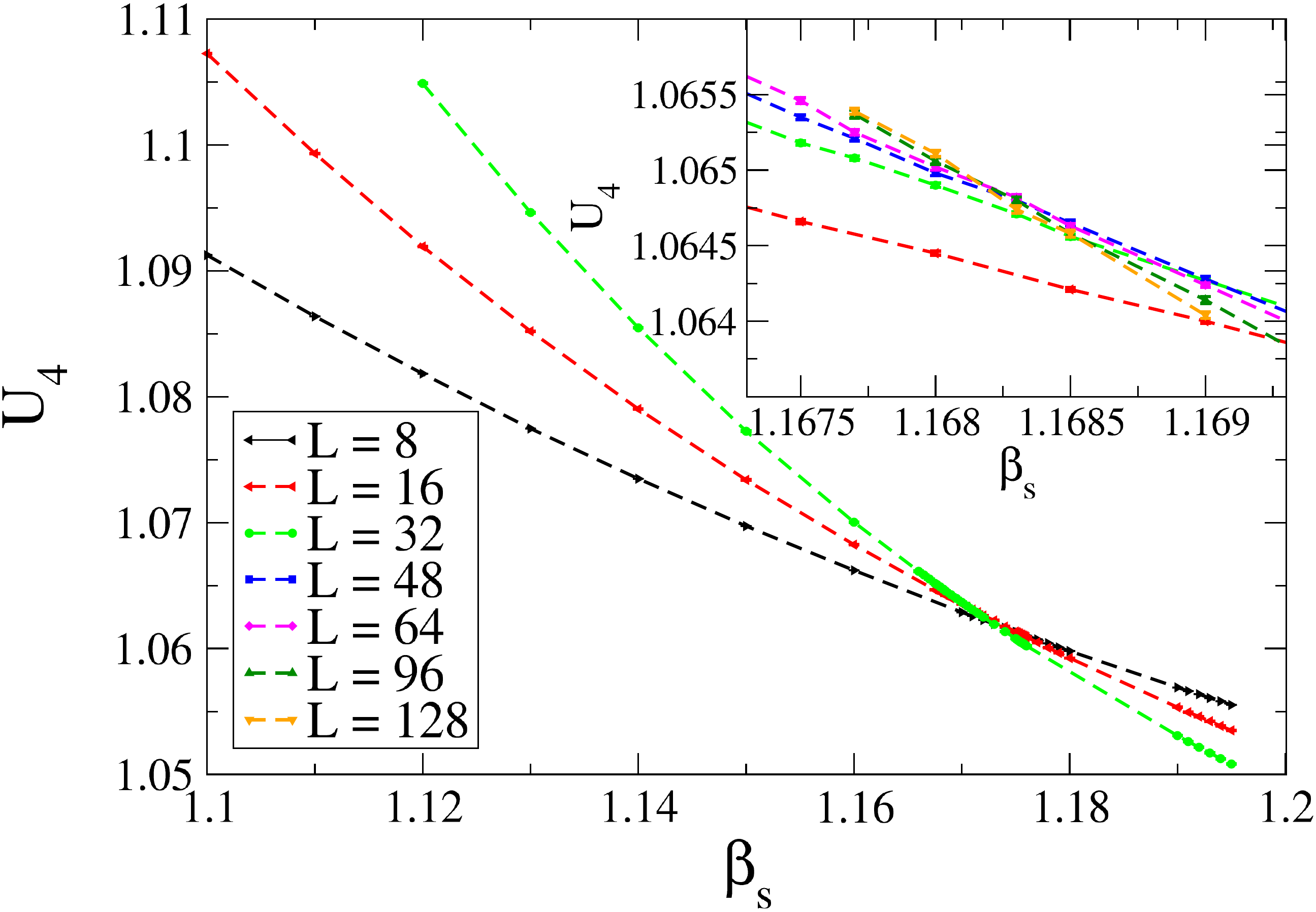}
    \caption{Surface Binder ratio $U_4$ at the special transition, as a function of $\beta_s$; Inset: MC data close to the special transition. Data are taken from Ref.~\cite{PT-20}.}
    \label{plot_U4}.
  \end{minipage}
  \hspace{1.5pc}
  \begin{minipage}{0.45\linewidth}
        \vspace{-2.3em}
    \centering
    \includegraphics[width=0.925\linewidth]{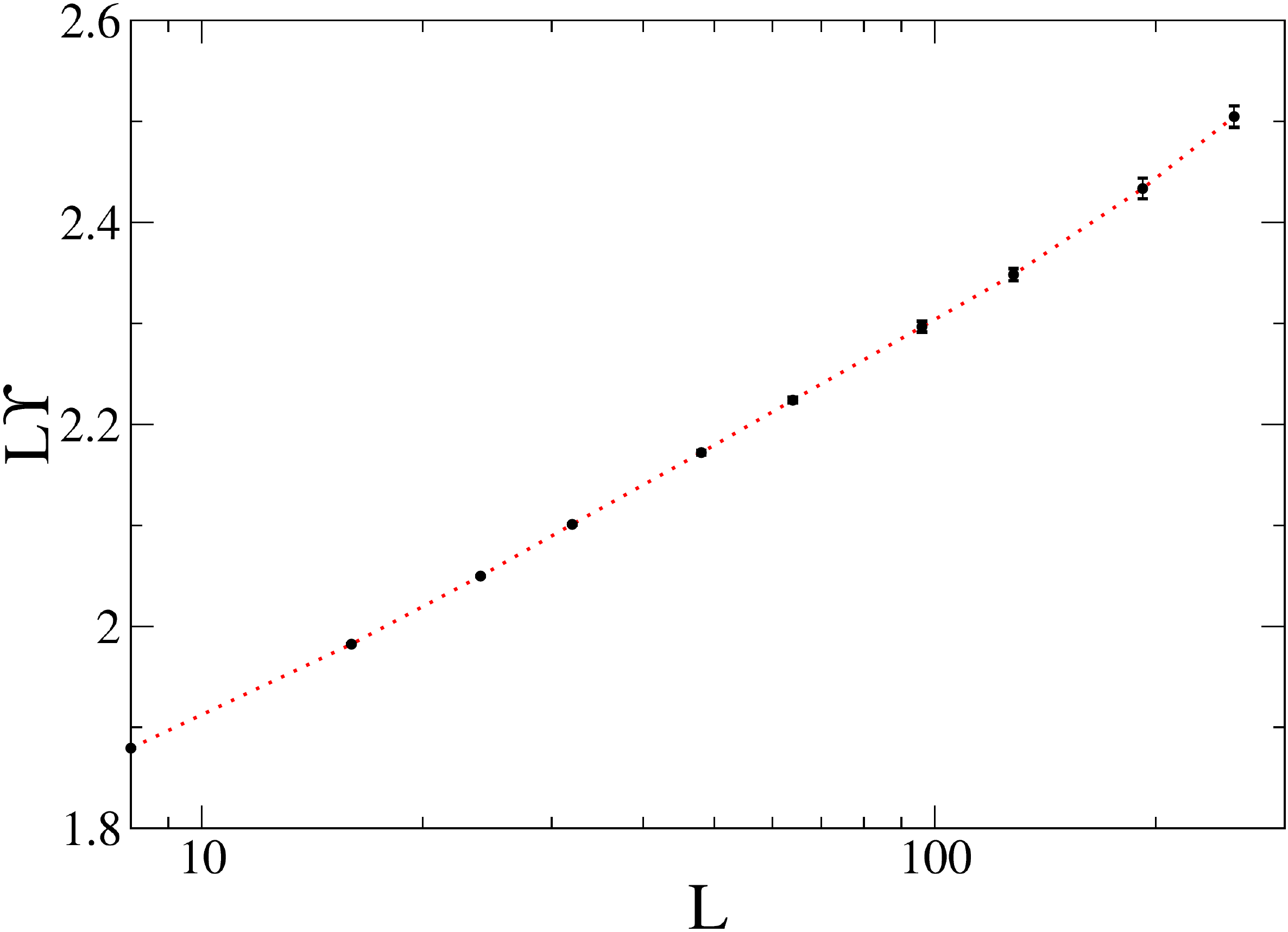}
    \vspace{0.5em}
    \caption{Helicity modulus at $\beta_s=1.5$, in the extraordinary phase, in a semilogarithmic scale. From Ref.~\cite{PT-20}.}
    \label{plot_helicity}
  \end{minipage}
\end{figure}
In Fig.~\ref{plot_U4} we show $U_4$ as a function of $\beta_s$. A scan over a wide range in $\beta_s$ reveals a crossing, indicative of a surface transition.
Close to the putative transition, we have sampled $U_4$ for lattice sizes up to $L=128$;
the data are shown in the inset of Fig.~\ref{plot_U4}.
We observe a rather slow increase of the slope of $U_4$ with $L$, such that a precision of $\approx 10^{-5}$ is needed in order to resolve the crossing.
A fit of $U_4$ to a suitable Taylor expansion of Eq.~(\ref{FSS_RGinv}), including also scaling corrections, allowed us to estimate the critical-point value $U_4^* \equiv U_4(\beta_{s,c}) = 1.0652(4)$.
We have used this value to analyze MC data, supplemented by additional simulations at $L=192$,
using FSS analysis at fixed RG-invariant \cite{Hasenbusch-99,HPTPV-07,PT-11}.
In this method, one fixes a chosen RG-invariant $R$ (here, $R=U_4$), thereby trading the statistical fluctuations of $R$ with fluctuations of a coupling constant driving the transition (here, $\beta_s$).
A discussion of the method can be found in Ref.~\cite{PT-11}.
To estimate the exponent $y_{\rm sp}$, we have computed
derivatives with respect to $\beta_s$ of RG-invariants $R=U_4$
and $R=Z_a/Z_p$, the ratio of the partition functions with antiperiodic and periodic BCs on a direction parallel to the surfaces; this ratio can be conveniently sampled with the boundary-flip algorithm \cite{Hasenbusch-93,CHPRV-01}.
At fixed $U_4$, $dR/d\beta_s$ behaves as $dR/d\beta_s \propto L^{y_{\rm sp}}$.
Suitable fits of $dU_4/d\beta_s$ and $d(Z_a/Z_p)/d\beta_s$ at fixed $U_4$, and as a function of $L$, including also scaling corrections, have delivered the estimate
\begin{equation}
  y_{\rm sp} = 0.36(1), \qquad \nu_{\rm sp}\equiv 1/y_{\rm sp} = 2.78(8).
  \label{ysp}
\end{equation}
Next, we have sampled the surface susceptibility $\chi_s$.
At fixed $U_4$ its leading scaling behavior is $\chi_s\propto L^{2-\eta_\parallel}$. Fits of $\chi_s$ resulted in the estimate
\begin{equation}
  \eta_\parallel = -0.473(2).
  \label{eta}
\end{equation}
FSS analysis at fixed $U_4$ has also allowed us to estimate the critical surface couplint at the onset of the special transition $\beta_{s,c}=1.1678(2)$.

\subsection{Extraordinary phase}
\label{sec:results:extraordinary}
The existence of a special transition implies that
for $\beta_s>\beta_{s,c}$ the surface displays an extraordinary phase.
To study it, we have simulated the model at $\beta_s=1.5$, for lattice sizes $8\le L \le 384$.
We have computed the helicity modulus $\Upsilon$ which
measures the response of the model to a torsion in the lateral BCs.
To compute it, one replaces the nearest-neighbor interaction along one lateral boundary in the Hamiltonian (\ref{model}) as
\begin{equation}
  \vec{\phi}_{\vec{x}}\cdot\vec{\phi}_{\vec{x}+\hat{e}_l} \rightarrow \vec{\phi}_{\vec{x}} R_{\alpha,\beta}(\theta)\vec{\phi}_{\vec{x}+\hat{e}_l},
  \label{torsion}
\end{equation}
with
 $\hat{e}_l$ the unit vector along one of the lateral directions, where periodic BCs are applied.
In Eq.~(\ref{torsion}) $R_{\alpha,\beta}(\theta)$ is a rotation matrix that rotates the $\alpha$ and $\beta$ components of $\vec{\phi}$ by an angle $\theta$.
For the present geometry, the helicity modulus is then defined as \cite{FBJ-73}
\begin{equation}
\Upsilon \equiv \frac{1}{L} \frac{\partial^2 F(\theta)}{\partial \theta^2}\Big|_{\theta=0},
\end{equation}
where $F$ is the total free energy.
In Fig.~\ref{plot_helicity} we show the product $\Upsilon L$, which exhibits a remarkable logarithmic growth, neither compatible with a standard critical phase, where $\Upsilon L\sim \text{const}$, nor with an ordered phase, where $\Upsilon \sim \text{const}$.
A logarithmic violation of FSS is also found in the ratio $\xi/L$ of the finite-size second-moment correlation length $\xi$ \footnote{See Appendix A of \cite{PTHAH-14} for a discussion on the definition of a finite-size correlation length.} over the size $L$.
Moreover, the two-point function on the surface exhibits a rather slow, but visible, decay.
These findings are indicative of the extraordinary-log scenario put forward in Ref.~\cite{Metlitski-20}.
In such a phase, the surface correlations decay as $C(x\rightarrow\infty)\propto \ln(x)^{-(N-1)/(2\pi\alpha)}$, where $N=3$ for the present case, and $\alpha$ is a universal RG parameter, determined by some amplitudes at the normal UC.
Furthermore, in the extraordinary-log phase a logarithmic violation of FSS is predicted, such that $\Upsilon L \simeq 2\alpha\ln(L)$ and $(\xi/L)^2\sim (\alpha/2)\ln L$ \cite{Metlitski-private}.
To further check this scenario, we have performed fits of various observables.
Fits of the surface two-point function provided an estimate $\alpha=0.15(2)$.
The quoted error bar has been estimated by comparing various fit results and should be taken with some caution, because in fitting the data we did not consider subleading corrections; these are potentially important, as found, e.g., in other critical models with marginal operators \cite{HPTPV-08b}.
Fits of $\xi/L$ to the expected logarithmic growth delivered $\alpha\approx 0.14$, consistent with the estimate coming from the correlations, although we observed some drift in the fitted value of $\alpha$ as a function of the minimum lattice size $L$ used in the fits.
Fits of $\Upsilon L$ gave less stable results, with $\alpha \gtrsim 0.11$.
All in all, despite the intrinsic difficulty in estimating a logarithmic exponent, we found a rough quantitative consistence of the scaling behavior with the scenario of an extraordinary-log phase.

\section{Summary}
\label{sec:summary}
In Ref.~\cite{PT-20} we have elucidated the boundary critical behavior of the three-dimensional $O(3)$ UC, in the presence of a 2D surface.
A FSS analysis of high-precision MC data allowed us to conclude that there is a special transition on the surface, in the presence of a critical bulk, and to compute the associated critical exponents.
The exponent $\eta_\parallel$ that we found is remarkably close to the nonordinary $\eta_\parallel$ exponent found in quantum MC studies of dimerized spin models \cite{SS-12,ZW-17,DZG-18,WPTW-18,WW-19,WW-20,ZDZG-20}.
This suggests that, for the geometrical settings where such a nonordinary exponent is found, these models are ``accidentally'' close to the special transition.
Unlike the ordinary and extraordinary surface phases, the special UC has a relevant non symmetry-breaking surface scaling field with dimension $y_{\rm sp}$ [see Eq.~(\ref{FSS_RGinv})].
Therefore, one generically needs a fine tuning of boundary interactions in order to realize the special UC.
Nevertheless, we notice that the value of $y_{\rm sp}$ [Eq.~(\ref{ysp})] is unusually small: this implies a slow crossover from the special fixed point when surface interactions are tuned away from the special transition.
Therefore, it may be possible to observe critical exponents similar to those of the special UC even if a model is not very close to the special transition.
This can provide an explanation to the observed nonordinary edge exponent $\eta_\parallel$ found in the aforementioned quantum spin models, without the need of a fine-tuning.
To further substantiate this hypothesis, it would be desirable to study in more detail the quantum-to-classical mapping \cite{Sachdev-book} of these spin models.

In the extraordinary phase, our MC data display slowly-decaying correlations and a remarkable logarithmic violation of FSS, which is indicative of the extraordinary-log phase scenario put forward in Ref.~\cite{Metlitski-20}.
Recently, this scenario has been put on a firmer ground in Ref.~\cite{PTM-21}, where we have computed the universal amplitudes of the normal UC that determine the onset and the value of the logarithmic exponent of the extraordinary-log phase.
We found a good agreement with MC simulations of the extraordinary phase presented in Refs.~\cite{PT-20}, \cite{HDL-21}, thus providing a nontrivial check of the connection between the normal and the extraordinary-log phases outlined in Ref.~\cite{Metlitski-20}.
A concurrent conformal bootstrap study \cite{PKMGM-21} found results in agreement with Ref.~\cite{PTM-21}.

\ack
FPT is funded by the Deutsche Forschungsgemeinschaft (DFG, German Research Foundation)--Project No. 414456783.
The author gratefully acknowledges the Gauss Centre for Supercomputing e.V. for funding this project by providing computing time through the John von Neumann Institute for Computing (NIC) on the GCS Supercomputer JUWELS at Jülich Supercomputing Centre (JSC) \cite{JUWELS}.

\section*{References}
\bibliography{francesco,extra}

\end{document}